\documentclass[twocolumn,showpacs,preprintnumbers,prd]{revtex4}
\usepackage{amsmath,amssymb}
\usepackage[dvips]{graphicx}
\usepackage{dcolumn}
\usepackage{bm}

\newcommand{\ua}{$\mathrm{U}_{\rm A}(1)$ }

\newcommand{\rmi}{\mathrm{i}}
\newcommand{\rmd}{\mathrm{d}}
\newcommand{\rme}{\mathrm{e}}
\newcommand{\gs}{g_{\mathrm{S}}}
\newcommand{\gd}{g_{\mathrm{D}}}

\begin{document}

\preprint{YITP-09-107}
\title{Model analysis on thermal UV-cutoff effects
 on the critical boundary in hot QCD}

\author{Jiunn-Wei Chen}
\email{jwc@phys.ntu.edu.tw}
\affiliation{Department of Physics
 and Center for Theoretical Sciences,
 National Taiwan University, Taipei 10617, Taiwan}%

\author{Kenji Fukushima}
\email{fuku@yukawa.kyoto-u.ac.jp}
\affiliation{Yukawa Institute for Theoretical Physics,
 Kyoto University, Kyoto 606-8502, Japan}

\author{Hiroaki Kohyama}
\email{kohyama@phys.sinica.edu.tw}
\affiliation{Institute of Physics, Academia Sinica,
 Taipei 115, Taiwan and\\
 Physics Division, National Center for Theoretical Sciences,
 Hsinchu 300, Taiwan}

\author{Kazuaki Ohnishi}
\email{kohnishi@phys.ntu.edu.tw}
 \altaffiliation{Present affiliation: GOGA, Inc., Tokyo.}
 \affiliation{%
 Department of Physics
 and Center for Theoretical Sciences,
 National Taiwan University, Taipei 10617, Taiwan }

\author{Udit Raha}
\email{udit.raha@unibas.ch}
 \affiliation{Department of Physics
 and Center for Theoretical Sciences,
 National Taiwan University, Taipei 10617, Taiwan }


\begin{abstract}
We study the critical boundary on the quark-mass plane associated with
the chiral phase transition in QCD at finite temperature.  We point
out that the critical boundaries obtained from the Lattice QCD
simulation and the chiral effective model are significantly different;
in the (Polyakov-loop coupled) Nambu--Jona-Lasinio (NJL) model we find
that the critical mass is about one order of magnitude smaller than
the value reported in the Lattice QCD case.  It is known in the
Lattice QCD study that the critical mass goes smaller in the continuum
limit along the temporal direction.  To investigate the temporal
UV-cutoff effects quantitatively we consider the (P)NJL model with
only finite $N_\tau$ Matsubara frequencies taken in the summation.  We
confirm that the critical mass in such a UV-tamed NJL model becomes
larger with decreasing $N_\tau$, which demonstrates a correct tendency
to explain the difference between the Lattice QCD and chiral model
results.
\end{abstract}

\maketitle

\textit{Introduction}~~
The phase diagram of quark matter as described by Quantum
Chromodynamics (QCD) is of great interest in theoretical and
experimental physics.  Since QCD has been established as the
fundamental theory of quarks and gluons, it is our ultimate goal to
understand all the phenomena related to the strong interactions from
the first principle theory.  It is difficult, however, to surpass
perturbative calculations due to largeness of the QCD coupling
constant in the low-energy regime.  One then has to make a choice
among non-perturbative techniques such as the Lattice QCD (LQCD)
simulation, some effective model descriptions sharing the same global
symmetry as QCD, etc. The Lattice QCD (LQCD) is a theoretical
framework of quarks and gluons in discretized
space-time~\cite{Wilson:1974sk} which is suitable for non-perturbative
investigations.  The Monte-Carlo simulations in the LQCD have been
recently developing to a reliable level not only for hadron spectrums
at zero temperature but also for thermal properties in hot QCD
matter~\cite{Philipsen:2008gf}.

\begin{figure}[tbh]
\begin{center}
\includegraphics[width=0.7\columnwidth]{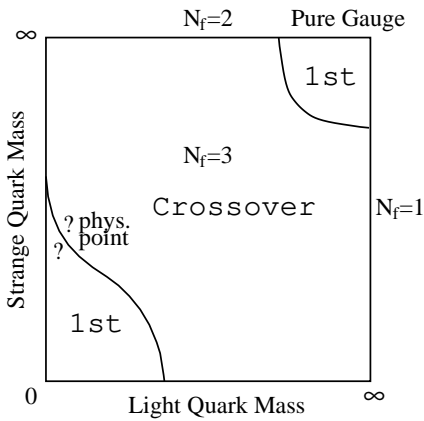}
\end{center}
\caption{Columbia plot at finite $T$ and $\mu=0$.}
\label{columbia}
\end{figure}

Critical phenomena in hot QCD can be studied in the LQCD
simulation~\cite{Ejiri:2009ac} and it is an important issue to
determine the order of the chiral phase transition as a function of
the temperature $T$ and the quark chemical potential $\mu$.  In this
paper, we study the critical boundary in the chiral phase transition
on the quark-mass plane, as represented by the ``Columbia
plot''~\cite{Brown:1990ev}.  Figure~\ref{columbia} shows a schematic
picture of the Columbia plot at $\mu=0$ where each point in the
parameter space of current quark masses $\{m_{u},\,m_{d},\,m_{s}\}$ is
marked by its order of the phase transition when $T$ is raised.
Usually one sets $m_{u}=m_{d} (\equiv m_{ud})$ and studies the order
of the phase transition on the $m_{ud}$-$m_{s}$ plane.  Symmetry
arguments show that it should be a first order phase transition in the
chiral limit ($m_{ud}=m_{s}=0$) \cite{Pisarski:1983ms}, while for
intermediate quark masses at which both chiral and center symmetries
are badly broken, only a smooth crossover is possible.  This defines
the second-order critical phase boundary that separates the
first-order and crossover regions.

 
The ``phys.\ point'' in Fig.~\ref{columbia} designates the ``real
world'' with physical current quark masses ($m_{ud}=5.5$ MeV and
$m_{s}=135.7$ MeV in a model study), whose location should crucially
decide the order of the phase transition at $\mu=0$.  More
importantly, the distance between the second-order critical boundary
and the physical point on the Columbia plot gives a rough measure of
how far the QCD critical point is separated from the $\mu=0$
situation.  Roughly speaking, the farther the physical point is
located away from the critical boundary, the larger $\mu$ is necessary
to make the critical surface hit on the physical point.

The critical boundary can also be analyzed by means of the LQCD
simulation since the Columbia plot is drawn at zero baryon density,
where the Binder cumulant is a useful quantity to identify the order
of the phase
transition~\cite{deForcrand:2006pv,deForcrand:2007rq,Endrodi:2007gc}.
To draw the Columbia plot we can use the three-flavor
Nambu--Jona-Lasinio (NJL) model~\cite{Fukushima:2008wg} which is a
model of quarks with the effective Lagrangian so constructed to have
the same global symmetry as
QCD~\cite{Klevansky:1992qe,Hatsuda:1994pi}.  (For other chiral model
studies see \cite{Kovacs:2006ym,Schaefer:2008hk}.)  Gauge symmetry is,
however, absent due to the absence of gluons in the model.  It is a
crucial aspect in the NJL model to incorporate chiral symmetry as an
important ingredient in understanding the properties of light
hadrons.  Further, the model can successfully describe the spontaneous
breaking of chiral symmetry at in the QCD vacuum and its restoration
at high temperature and/or density.  The three-flavor NJL model gives
a first order phase transition in the chiral limit by virtue of the
\ua symmetry breaking term~\cite{Pisarski:1983ms}, while, at
intermediate quark mass, the chiral restoration results in a crossover
due to explicit chiral symmetry breaking by finite quark mass as long
as $\mu\simeq 0$.  Thus, the NJL model should be able to reproduce the
expected structure of the Columbia plot (i.e.\ critical boundary in
the left-bottom corner in Fig.~\ref{columbia}), which can be compared
with the LQCD results.

Although the NJL model successfully describes the qualitative feature
of the chiral phase transition, there remains a significant difference
between the critical boundaries from the NJL model and the LQCD
simulations.  In fact, the value of the critical quark mass is about
one order of magnitude different, which is crucial in the QCD critical
point search.  One may have seen a scatter plot of the critical point
locations from the LQCD attempts as well as the chiral model
approaches as shown in \cite{Stephanov:2004wx}.  However, there is a
huge discrepancy in the critical mass boundary already at zero baryon
density depending on the adopted methods, and then, how can one trust
any of them at finite density?  Let us pick some concrete numbers up.
We define the critical quark mass $m_{c}$ as a point where the
$m_{ud}=m_{s}$ symmetric line intersects with the critical boundary.
Then, $m_c$ is around $14$ MeV in the LQCD
simulation~\cite{deForcrand:2006pv} but it is only $1.1$ MeV from the
NJL model.

In the LQCD studies~\cite{deForcrand:2007rq,Endrodi:2007gc}, it was
argued that the first-order region drastically shrinks by removal of
the discretization errors.  In \cite{Endrodi:2007gc}, for example, the
authors used the Symanzik improved gauge and stout improved fermionic
action and concluded that the critical mass goes below $7\%$ (and
$12\%$) of the physical quark mass on the $N_\tau=4$ (and $N_\tau=6$
respectively) lattice.  The conclusion in \cite{deForcrand:2007rq} is
consistent with this qualitatively though the quantitative change is
not such drastic.  Because the number of spatial points is large
enough (16, 24, etc) in the LQCD simulations, smallness of $N_\tau$ in
the temporal direction is crucial.  In other words, the temporal UV
cutoff $\pi N_\tau T$ has drastic effects on the critical mass
boundary even at zero density (and hence on the QCD critical point at
finite density as well, which is beyond our current scope in this
paper).

Here, we shall examine the chiral critical boundary by imposing a
finite UV cutoff in the temporal (thermal) direction in the NJL model
to mimic the LQCD situation.  In this way, we can give a quantitative
prediction about how much the critical boundary moves by the finite
$N_\tau$ effects.  In terms of Fourier transformed variables the
summation over finite-$N_\tau$ temporal sites amounts to the Matsubara
summation over finite-$N_\tau$ frequencies.  Hereafter, throughout
this paper, we will drop the subscript and simply write as $N$ for
notational simplicity.  In what follows, we present a brief description
of this modified NJL model and discuss the machinery of how to deal
with finite $N$ in this model treatment.  Later, we shall also present
some results using the Polyakov-loop coupled NJL (PNJL)
model~\cite{Fukushima:2008wg,Fukushima:2003fw,Ratti:2005jh,Fu:2007xc,Ciminale:2007sr},
for the PNJL model is a much better description of QCD thermodynamics
near the critical temperature.
\\

\textit{Modified NJL model}~~ 
The three-flavor NJL model Lagrangian is,
\begin{align}
 \mathcal{L}_{\mathrm{NJL}} &= \bar{q}\left( \rmi\partial\!\!\!/
  - \hat{m}\right)q + \mathcal{L}_4 + \mathcal{L}_6 ,
\label{LNJL} \\
 \mathcal{L}_4 &= \frac{\gs}{2} \sum_{a=0}^8 \left[
  \left( \bar{q}\lambda_a q\right)^2
  + \left( \bar{q}\,\mathrm{i}\gamma_5 \lambda_a q \right)^2
  \right] ,
\label{L_4} \\
 \mathcal{L}_6 &= -\gd \left[ \det\bar{q}_i (1-\gamma_5) q_j 
 +\text{h.c.\ } \right] .
\label{L_6}
\end{align}
The current quark mass matrix $\hat{m}$ in the kinetic term takes the
diagonal form $\hat{m}=(m_{u},m_{d},m_{s})$, and we set
$m_{u}=m_{d}\;(\equiv m_{ud})$.  In the above $\mathcal{L}_4$
expresses the four-fermion contact interaction with the coupling
constant $\gs$, where $\lambda_a$ is the Gell-Mann matrix in flavor
space, and $\mathcal{L}_6$ is the Kobayashi-Maskawa-'t~Hooft
interaction~\cite{Kobayashi:1970ji,Kobayashi:1971qz,'tHooft:1976up}
with the coupling strength $\gd$.  The determinant runs over the
flavor indices, which leads to six-fermion interaction terms in the
three-flavor case.  This interaction explicitly breaks the \ua
symmetry, whose microscopic origin comes from
instantons~\cite{'tHooft:1976up}.

We should solve the gap equations which we can get by differentiating
the thermodynamic potential $\Omega$ with respect to the constituent
quark masses;
\begin{equation}
 \frac{\partial\, \Omega}{\partial\, m_u^*}=0 \, , \qquad
 \frac{\partial\, \Omega}{\partial\, m_s^*}=0.
\end{equation}
Here $m_u^*$ and $m_s^*$ are the constituent masses for $u$ and $s$
quarks.  The thermodynamic potential is defined by
$\Omega = -\ln Z /(\beta V)$, with the partition function $Z$, the
inverse temperature $\beta=1/T$, and the volume of the system $V$.

In the mean-field approximation for the NJL Lagrangian~\eqref{LNJL} we
obtain the following gap equations,
\begin{equation}
 \begin{split}
  m_u^* &= m_u + 2\rmi \,\gs\, N_c \mathrm{tr}S^u
   - 2\gd N_c^2 (\mathrm{tr}S^u)(\mathrm{tr}S^s) , \\
  m_s^* &= m_s + 2\rmi \,\gs\, N_c \mathrm{tr}S^s
   - 2\gd N_c^2 (\mathrm{tr}S^u)^2 .
\end{split}
\label{gap}
\end{equation}
Here $N_c=3$ is the number of colors and $\mathrm{tr}S^i$ is the
chiral condensate given explicitly by
\begin{equation}
 \rmi\; \mathrm{tr}S^i = 4m_i^* \int \! \frac{d^3p}{(2\pi)^3}\;
  (\rmi\,T)\!\! \sum_{n=-\infty}^{\infty}
  \frac{\rmi}{(\rmi\,\omega_n)^2 - E_i^2} \, ,
\label{trace}
\end{equation}
where the Matsubara frequency is
$\omega_n=(2n+1)\pi T$ and the quasi-particle energy is
$E_i= \sqrt{\bm{p}^2 + m_i^{*\,2}}$.  A detailed derivation of the gap
equations is clearly given in \cite{Klevansky:1992qe,Hatsuda:1994pi}.

To study the effect of the finite cutoff in the temporal direction on
the critical boundary, we then perform the following modification in
the Matsubara frequency sum;
\begin{equation}
 \sum_{n=-\infty}^\infty \longrightarrow \sum_{n=-N+1}^{N} .
\end{equation}
We shall reveal the critical boundary in the $m_{ud}$-$m_{s}$ plane
for various values of $N$.  To this end, we adjust the model
parameters to fit the physical quantities for a given $N$ (each
parameter set representing a modified NJL model with a different
temporal cutoff).  This is reminiscent of the familiar Wilsonian
renormalization group (RG) flow.  After the high frequency modes are
integrated out, in principle, the coefficients of the operators run on
the RG flow.  Here we remark that we neglect interaction operators in
other channels than those in Eqs.~\eqref{LNJL}-\eqref{L_6}) since
their effects are small in the mean-field calculations performed
below (see e.g., \cite{Klevansky:1992qe} and refs. therein).

The following are the model parameters in the three-flavor NJL model;
\begin{center}
 \begin{tabular}{l@{\hspace{0.5cm}}l}
  current quark masses & $m_{ud}$, $m_s$ \;,\\
  three-momentum cutoff & $\Lambda$ \;,\\
  four-point coupling constant & $\gs$ \;,\\
  six-point coupling constant & $\gd$ \;.
 \end{tabular}
\end{center}

\noindent
As for the light current quark masses, we fix $m_{ud}=5.5$ MeV
following \cite{Hatsuda:1994pi}, which gives isospin symmetric
results and thus $m_u^*=m_d^*$.  The remaining parameters are
determined by a fit to the four physical quantities: the pion mass $m_\pi$,
the pion decay constant $f_\pi$, the kaon mass $m_K$, and the
$\eta^{\prime}$ mass $m_{\eta^{\prime}}$, whose empirical values are
\begin{align}
 &m_{\pi}=138 \text{ MeV}\,, && f_{\pi}=93 \text{ MeV}\,, \notag\\
 &m_K=495.7 \text{ MeV}\,, && m_{\eta^{\prime}}=958 \text{ MeV} \,. 
\end{align}


\begin{table}[tbp]
 \begin{tabular}{c@{\hspace{0.5cm}}c@{\hspace{0.5cm}}c%
@{\hspace{0.5cm}}c@{\hspace{0.5cm}}c}
 \hline\hline
$N$ & $m_s$ (MeV) & $\Lambda$ (MeV) & $\gs\Lambda^2$ & $\gd\Lambda^5$ \\
 \hline
$15$ & $134.7$ & $631.4$ & $4.16$ & $12.51$ \\ 
$20$ & $135.0$ & $631.4$ & $4.02$ & $11.56$ \\ 
$50$ & $135.3$ & $631.4$ & $3.82$ & $10.14$ \\ 
$100$ & $135.4$ & $631.4$ & $3.75$ & $9.69$ \\ 
$\infty$ & $135.7$ & $631.4$ & $3.67$ & $9.29$ \\
 \hline\hline
\end{tabular}
\caption{Fitted parameters for various values of $N$.}
\label{parameters}
\end{table}


\textit{Numerical results}~~
The limit of $N=\infty$ invariably corresponds to the standard
framework of the NJL model and we then use the same parameter set fixed in
\cite{Hatsuda:1994pi}.  For finite $N$, we have picked up a temperature
$T=50$ MeV at which we have carried out our fitting procedure.  
This choice is just
because of technical convenience.  Besides, calculations at $T=50$ MeV
should be sufficiently close to those at zero temperature since the
constituent quark mass is then several times larger than the
temperature.  The fitted parameters in this way for $N=15$, $20$,
$50$, $100$ at $T=50$ MeV are displayed in Tab.~\ref{parameters}.


\begin{figure}[tbh]
\begin{center}
\includegraphics[width=0.9\columnwidth]{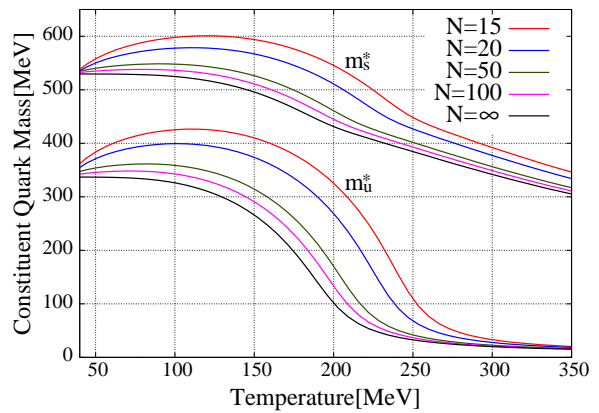}
\end{center}
\caption{Constituent quark masses $m_u^*$ (lower) and $m_s^*$ (upper)
  as a function of $T$ for various values of $N$.}
\label{constituent}
\end{figure}


As seen from the table, the couplings $\gs$ and $\gd$ become larger
with decreasing $N$.  If we fixed the coupling constants, the effect
of smaller $N$ would be to reduce the constituent quark masses.
Therefore, to keep $m_i^*$ unchanged, we need larger $\gs$ and $\gd$
to compensate for this reduction (see Eq.~(\ref{gap})).  Note that, as
expected, the whole behavior of $m_u^\ast$, $m_s^\ast$ as a function of
$T$ monotonically approach the standard (i.e.\ $N=\infty$) case
with increasing $N$.  In Fig.~\ref{constituent}, we only show the
numerical results above $T=40$ MeV because the small-$T$ region is
badly affected by finite $N$ and the calculations are no longer
reliable there.  We can understand this intuitively;  $1/T$ is the
extent along the imaginary-time (thermal) direction and thus, the
extent becomes longer as $T$ goes smaller.  Hence, at lower $T$,
reliable estimates need more $N$.  This means, fixing the value of $N$,
the choice of the small-$T$ regions would automatically lead to unphysical 
results.


\begin{figure}[tbh]
\begin{center}
\includegraphics[width=1.0\columnwidth]{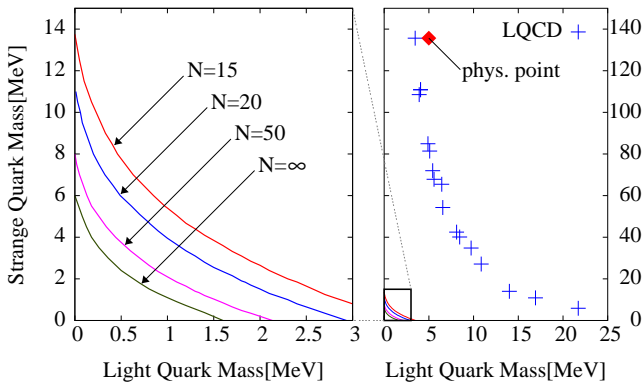}
\end{center}
\caption{Critical boundaries on the $m_{ud}$-$m_s$ plane in the NJL
  model (left panel) as compared with the LQCD results with $N_\tau=4$
  (right panel).}
\label{us_plane}
\end{figure}


In Fig.~\ref{us_plane}, we display the result for the critical boundary
in our modified NJL model with the finite Matsubara frequency
summations: $N=15$, $20$, $50$, $\infty$, along with the results from
the LQCD simulation with $N_\tau=4$~\cite{deForcrand:2006pv} for
comparison.  We find that the first-order region widens as $N$
decreases.  This observation qualitatively agrees with the arguments
in the LQCD simulations that the first-order region tends to get
bigger than its genuine size due to the finite UV-cutoff effects.
Quantitatively, however, the critical boundaries still look far from
the LQCD results at $N_\tau=4$.  We comment that not only the temporal
UV-cutoff but also the spatial (three-momentum) cutoff leads to
similar alteration in the critical boundary i.e.,  we have found that
lowering $\Lambda$ also pushes the critical mass up, though such
effects are only minor.
\\

\textit{The PNJL model extension}~~
It is also worth studying the critical boundary using the PNJL
model~\cite{Fukushima:2008wg,Fukushima:2003fw,Ratti:2005jh,Fu:2007xc,Ciminale:2007sr}.
This is because, in view of the results in
\cite{Roessner:2006xn,GomezDumm:2008sk}, the location of the critical
point is significantly moved toward higher $T$ by the inclusion of the
Polyakov loop which suppresses unphysical quark excitations.
Furthermore, the PNJL model is quite successful to reproduce the LQCD
observables for QCD thermodynamics such as the quark number
susceptibility, the interaction measure, the sound velocity, etc.
Because the PNJL model nicely grasps the essential nature of the QCD
phase transitions, it would be rather mysterious if only the critical
boundary on the $m_{ud}$-$m_s$ shows a huge discrepancy.
 
The model is defined by the following
Lagrangian~\cite{Fukushima:2003fw,Ratti:2005jh},
\begin{align}
 & \mathcal{L}_{\mathrm{PNJL}} = \mathcal{L}_0+\mathcal{L}_4
  +\mathcal{L}_6 + \mathcal{U}(\Phi,\Phi^*,T) \, , \\
 & \mathcal{L}_0 = \bar{q}\left(\rmi \partial \!\!\!/
  - \rmi\gamma_4 A_4 - \hat{m}\right)q \, , \\
 & \mathcal{U}(\Phi,\Phi^*,T) = -bT\bigl\{ 54\,\rme^{-a/T}\Phi\Phi^*
  \notag \\
 &\qquad +\ln\bigl[1 - 6\Phi\Phi^* + 4(\Phi^3+\Phi^{*\,3})
  - 3(\Phi\Phi^*)^2 \bigr] \bigr\}\, ,
\end{align}
where $\mathcal{U}$ is the effective potential in terms of the
Polyakov loop.  Here $\Phi$ and $\Phi^*$ are the traced Polyakov loop
and the anti-Polyakov loop which are order parameters for
deconfinement.  They are defined by $\Phi=(1/N_c)\mathrm{tr}L$,
$\Phi^*=(1/N_c)\mathrm{tr}L^\dagger$ with
$L=\mathcal{P} \exp[\rmi\int_0^\beta \!\rmd\tau A_4]$.  We set the
parameters $a$ and $b$ as $a=664$ MeV, $b\!\cdot\!\Lambda^{-3}=0.03$
following \cite{Fukushima:2008wg}.


\begin{figure}[tbh]
\begin{center}
\includegraphics[width=1.0\columnwidth]{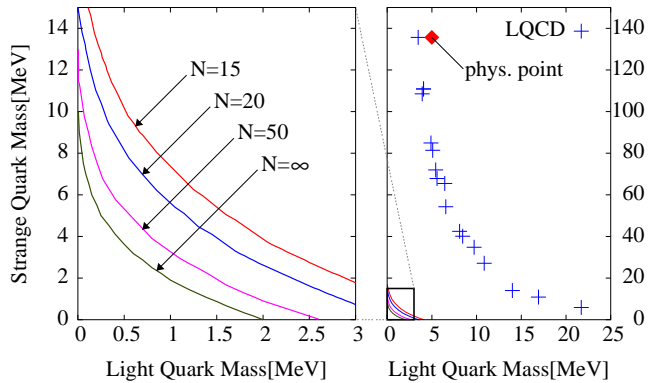}
\end{center}
\caption{Critical boundaries on the $m_{ud}$-$m_s$ plane in the PNJL
  model (left panel) as compared with the LQCD results with $N_\tau=4$
  (right panel).}
\label{us_plane_P}
\end{figure}


Because the Polyakov-loop part is relevant only at finite $T$ (as high
as comparable with $T_c$), we can simply employ the same parameter
sets, as fixed previously in the NJL model case with finite $N$ (see
Tab.~\ref{parameters}).  Then, we show the results for the critical
boundary determined by the same procedure in Fig.~\ref{us_plane_P}.
Here again, we see expanding behavior of the first-order region with
decreasing $N$.  Although the PNJL results are quantitatively
different from the NJL ones, the qualitative tendency is the same and,
besides, the difference is not really substantial.  The fact that the
PNJL model results are such close to the NJL ones is in fact
non-trivial.  The Polyakov loop generally governs the thermal
excitation of quarks near $T_c$ and particularly unphysical quarks
below $T_c$ are prohibited by small $\Phi$ and $\Phi^\ast$.  This
leads to the consequence that the critical point temperature is lifted
up to twice as high.  From the comparison between Figs.~\ref{us_plane}
and \ref{us_plane_P}, gives us confidence that the smallness of the
first-order region is {\em not} simply an NJL-model artifact with 
unphysical quark excitations.
\\

\textit{Concluding remarks}~~
Our model analysis yields intriguing results that the critical masses
become larger with truncation in the Matsubara frequency summation.
This effect is not large enough to give a full account for the
quantitative difference in the critical boundaries between the LQCD
and NJL model studies.  Nevertheless, it raises an interesting
possibility that the critical masses in the LQCD simulation
become smaller if the LQCD simulation accommodates larger $N_\tau$.
This is an important message since the LQCD simulation at $N_\tau=4$,
$6$, and even $8$ may have a potential danger that the LQCD simulation
overestimates the strength of the first-order phase transition, and
thus the critical point emerges at small density due to lattice
artifacts.

An interesting extension of our study is to determine the chiral phase
transition boundary on the plane with $T$ and the baryon (quark)
chemical potential $\mu$.  It is a common folklore that the QCD
critical point should appear at some $T_E$ and $\mu_E$.  However, it
is highly non-trivial to determine the order of the phase transition
at non-zero $\mu$.  Recent LQCD simulations at small $\mu$
(i.e.\ expansion as a power of $\mu/T$) disfavor the existence of the
critical point in the QCD phase
diagram~\cite{deForcrand:2006pv,deForcrand:2007rq}.  There have been
theoretical efforts to understand this LQCD finding in the context of
the chiral effective model
approaches~\cite{Fukushima:2008is,Chen:2009gv}.  However, the errors
of the LQCD simulations grow bigger at higher $\mu$ and, to be worse,
the (P)NJL-model analyses can be trusted only at a qualitative level
so far once $\mu$ gets large.  It is still inconclusive, therefore,
whether the critical point actually exists or not.  Our model analysis shows
that increasing $N_\tau$ leads to smaller critical quark masses.  
This result might indicate to further minify the possibility for finding 
the QCD critical point using the LQCD simulation unless the 
discretization errors are under good theoretical control.

JWC, KO and UR are supported by the NSC and NCTS of Taiwan.  
KF is supported in part by Japanese MEXT grant No.\ 20740134 and 
the Yukawa International Program for Quark Hadron Sciences.  
UR also thanks INT, Seattle for the kind hospitality during the course
of this work. 



\end{document}